\documentclass[prb]{revtex4} 

\usepackage{amsmath}

\newcommand{\beq}{\begin{equation}}
\newcommand{\eeq}{\end{equation}}

\begin{document}

\title{Evolution and the second law of thermodynamics}
\author{Emory F. Bunn} 
\email{ebunn@richmond.edu}
\affiliation{Department of Physics, University of Richmond, Richmond, Virginia 23173}

%\date{\today}

\begin{abstract}
Skeptics of biological evolution often claim that evolution requires
a decrease in entropy, giving rise to a conflict
with the second law of thermodynamics. This argument 
is fallacious because it neglects the large increase in entropy
provided by sunlight striking the Earth. A recent article
provided a quantitative assessment of the
entropies involved and showed explicitly that there
is no conflict. That article rests on an unjustified assumption
about the amount of entropy reduction involved in evolution. I
present a refinement of the argument that does not rely on this
assumption.
\end{abstract}

\maketitle

\section{Introduction}
\label{sec:intro}
Daniel Styer recently addressed the
claim that evolution requires 
a decrease in entropy and therefore is in
conflict with the second law of thermodynamics.\cite{styer}
He correctly explained that 
this claim rests on misunderstandings about the nature of
entropy and the second law. The second
law states that the total entropy of a closed system must never
decrease. However, the Earth is not a closed system and is constantly
absorbing sunlight,
resulting in an enormous increase in entropy, which can counteract
the decrease presumed to be required for evolution.
This argument is known to those who defend evolution in 
evolution-creationism debates,\cite{talkorigins} but 
it is usually described in a general, qualitative 
way. 
Reference~\onlinecite{styer}
filled this gap with a quantitative argument.

In the following I present a more
robust quantitative argument. We begin
by identifying the appropriate closed system to which
to apply the second law. We find that the second law requires that
the rate of entropy increase
due to the Earth's absorption of 
sunlight, $(dS/dt)_{\rm sun}$, must be sufficient
to account for the rate of entropy
decrease
required for the evolution of life, $(dS/dt)_{\rm life}$ (a negative quantity).
As long as
\beq
\left(dS\over dt\right)_{\rm sun}+\left(dS\over dt\right)_{\rm life}\ge 0,
\label{eq:secondlaw}
\eeq
there is no conflict between
evolution and the second law.

Styer estimated both $(dS/dt)_{\rm sun}$ and $(dS/dt)_{\rm life}$ 
to show
that the inequality (\ref{eq:secondlaw}) is satisfied, but his argument
rests on an unjustified and probably incorrect assumption
about $(dS/dt)_{\rm life}$.\cite{styer}
I will present a modified version of the argument which does not depend
on this assumption and which 
shows that the entropy decrease required for evolution is orders of magnitude
too small to conflict with the second law of thermodynamics.

\section{Entropy provided by Sunlight}
\label{sec:sunlight}

Let us begin by justifying the inequality (\ref{eq:secondlaw}).
The Earth maintains an approximately constant temperature by absorbing
energy from the Sun and radiating energy at an almost equal
rate. To consider the application of the second
law of thermodynamics to these processes we first identify
a closed system that is large enough that these energy flows may be 
considered to be internal to the system. Let us take our system
to be the Earth, the Sun, and the outgoing thermal radiation
emitted by these bodies. We will ignore interactions of this
radiation with
bodies other than Earth and Sun and consider the outgoing radiation
from each to form an ever-expanding spherical halo.
In this system no entropy is produced by emission
of radiation from the Sun, because this process is a flow
of energy from the Sun to its radiation field at the same 
temperature. The same applies to radiation emitted by the
Earth. Entropy production occurs only when radiation from
the Sun is absorbed on the Earth, because this absorption
represents energy flow between parts of the system at different
temperatures. 

Let 
$T_\odot$ and $T_\oplus$ be the temperatures of
Sun and Earth respectively, and let $P$ be the solar power absorbed
by Earth. (To be precise, $P$ is the net 
flow from 
Sun to Earth,
including the backward flow of energy from Earthshine being absorbed
on the Sun, but the latter contribution is negligible.)
The Earth gains entropy at the rate $P/T_\oplus$, and the Sun's radiation field 
loses entropy at the rate $-P/T_\odot$. 
The rate of entropy production is
\beq
\left(dS\over dt\right)_{\rm sun}={P\over T_\oplus}-
{P\over T_\odot}\approx {P\over T_\oplus},
\eeq
where the last approximate equality reflects 
the fact that the Sun's temperature is much larger than the Earth's.
If we assume that the evolution of life requires the entropy to decrease at the rate $(dS/dt)_{\rm life}$, the second law
of thermodynamics applied to this system gives Eq.~(\ref{eq:secondlaw}).

By using values for the solar constant\cite{labs} and Earth's 
albedo,\cite{goode} Styer\cite{styer} found that Earth absorbs 
solar radiation at a rate of $P=1.2\times 10^{17}$\,W. If we use
$T_\oplus=300$\,K as a rough estimate of Earth's
temperature,
we find that
\beq
\left(dS\over dt\right)_{\rm sun}={P\over T_\oplus}=4\times 10^{14}\,{\rm (J/K)/s}=
(3\times 10^{37}k)\,{\rm s}^{-1},
\label{eq:ssun}
\eeq
where $k$ is Boltzmann's constant.

In this estimate we did not include any entropy increase
due to thermalization of the radiant energy emitted by the Earth.
If we assume that this radiation eventually
thermalizes with the cosmic background (CMB) radiation in deep space, then an additional, much larger entropy increase results:
$(dS/dt)_{\rm CMB}=P/T_{\rm CMB}=4\times 10^{16}$\,(J/K)/s.
We may not include this entropy production
in accounting for evolution, though. One reason is that this thermalization
probably never occurs: the mean free path of a photon in intergalactic space
is larger than the observable Universe and is probably
infinite.\cite{kt} In any case, even if thermalization does occur,
it happens far in the future and at great distances
from Earth and so is not available to drive evolution on Earth.
For this reason we may ignore the existence of distant
thermalizing matter in defining the system to which we 
apply the second law.
The argument in Sec.~\ref{sec:myestimate}, 
which concludes that inequality
(\ref{eq:secondlaw}) is satisfied, would only be strengthened
if this extra entropy were included.\cite{endnote1}

\section{Evolutionary decrease in entropy}
\label{sec:styer}
We now consider $(dS/dt)_{\rm life}$.
Styer's argument 
relies on
the assumption:
``Suppose that, due to evolution, each individual organism is 1000 times
`more improbable' than the corresponding individual was 100
years ago. In other words, if $\Omega_i$ is the number of microstates
consistent with the specification of an organism 100 years ago, and 
$\Omega_f$
is the number of microstates consistent with the specification of
today's `improved and less probable' organism, then 
$\Omega_f =
10^{-3}\Omega_i$. 
I regard this as a very generous rate of evolution, but you
may make your own assumption.''\cite{styer}

The fact that no justification is provided for this assumption
undermines the persuasive power of the argument. 
Moreover, 
far from being generous, a probability ratio of $\Omega_i/\Omega_f=10^3$ 
is probably much
too low. One of the central ideas of statistical mechanics is that
even tiny changes in a macroscopic object
(say, one as large as a cell) result in exponentially large changes
in the multiplicity (that is, the number of accessible microstates). 

I will illustrate this idea by some order of magnitude estimates.
First, let us address the precise meaning of the phrase ``due
to evolution.'' If a child grows up to be 
slightly larger than her mother
due to improved nutrition, we do not describe this change as due to
evolution, and thus we might not count the associated
multiplicity reduction in the factor $\Omega_i/\Omega_f$. 
Instead we might count only 
changes such as the turning on of a new gene as being due 
to evolution. 
However, this narrow view would be incorrect. 
For this argument we
should do our accounting in such a way that all biological
changes are included. Even if a change like the increased
size of an organism is not the direct result of evolution for
this organism in this particular generation, 
it is still ultimately due to evolution in the
broad sense that all life is due to evolution.
All of the extra proteins, DNA molecules, and other complex structures 
that are present
in the child 
are there because of evolution at some point 
in the past if not in the present, and they
should be accounted for in our calculation.

To see that this broad sense of evolution is 
the correct one to apply,
consider a thought experiment.  Suppose
that the entropy reduction due to life in this broad sense were
computed and found to be greater than the entropy provided by
sunlight.  Creationists would justifiably cite this finding
as proof that
evolution is in conflict with the second 
law.  

We now make some estimates of the required
multiplicity reduction.
We consider the case of an {\it E.\ coli}
bacterium. We will first consider the reduction in multiplicity associated with building this organism from
scratch. Following Styer,\cite{styer} we will then imagine a series of
ever-simpler 
ancestors of this organism at 100 year intervals, stretching back over
the four billion year history of evolution. Each organism
is somewhat more improbable than its ancestor from the previous century,
and the product of all of these multiplicity reductions must
be sufficient to account for the total required multiplicity 
reduction.

The entropy reduction associated with the evolution of life comes
in many forms, consisting of the construction of complex
structures from simpler building blocks. For simplicity, we will
consider just one portion of this process, namely the construction
of proteins from their constituent amino acids. Because we will neglect
other processes (such as the synthesis
of the amino acids in the first place and the formation of other macromolecules), we will greatly underestimate the required multiplicity 
reduction.

An
{\it E.\ coli} bacterium has about $4\times 10^6$ 
protein molecules.\cite{endnote2}
(This number refers not to the number of distinct types of
protein, but to the total number of protein molecules in the 
cell.)
We will find the multiplicity cost of assembling all of these molecules
by first considering the multiplicity cost of assembling
a single protein molecule.
Imagining assembling
the protein one amino acid at a time. At each step we must take
an amino acid that was freely moving through the cell and place it in
a specific position relative to the others that have already been assembled.
If the amino acids were previously in a dilute
solution in the cell, then the multiplicity loss due to 
each such step is approximately $n_Q/n$, where $n$ is the number density
of amino acids and $n_Q$ is the density at which the amino acids
would reach quantum degeneracy.\cite{endnote3}
This ratio is large: 
amino acids in a cell are far from degenerate. To assemble
a protein with $N_a$ amino acids, we would repeat this process
$N_a-1$ times, resulting in the large number
$\Omega_i/\Omega_f\sim (n_Q/n)^{N_a-1}$. For instance, if $n_Q/n=10$ (much too low) and $N_a=300$ (about the average size of a 
protein\cite{endnote2}),
the multiplicity ratio is $\sim 10^{299}$ for
the production of a single protein molecule.

If we use this conservative estimate for the multiplicity change
associated with the formation of a single protein molecule, we 
estimate the multiplicity reduction required
to assemble all of the proteins in the bacterium to be
$\sim (10^{299})^{4\times 10^6}\sim 10^{10^9}$.
If the entire 4 billion years ($4\times 10^7$ centuries) 
of biological
evolution were required to achieve this number, we would require
a multiplicity reduction of $(10^{10^9})^{1/(4\times 10^7)}=
10^{25}$ in each century, not
$10^3$.

These estimates are extremely rough. For example, they neglect
the internal degrees of freedom of the protein (which are 
far fewer than those of the free amino acids), and 
entropy changes due to the energy absorbed or emitted during the formation
of chemical bonds. To include the latter we note that the
multiplicity change associated with a chemical reaction is $e^{\mu/kT}$.
The chemical potential $\mu$ in a chemical reaction
is of order 1\,eV ($\sim 10^{-19}$\,J) or more, 
implying multiplicity changes of order $e^{40}\approx 10^{17}$ 
for each chemical bond formed or broken
at biological temperatures. Because hundreds of chemical bonds must be
formed in assembling each protein molecule, 
the resulting factor will again be 
exponentially large.
This sort of number is the ante to enter this particular 
game.

Because bacteria appeared very early in
evolution, we should assume a time period much shorter
than four billion years, and
hence a still larger multiplicity reduction per century would be required.
Similarly, if we considered more than just the formation
of proteins, or if we considered a large multicellular
organism, the required factor would be much greater.

Rough as these arguments are, they establish that there
is reason to doubt the factor $10^3$ that plays an important
role in Styer's argument, rendering his argument 
unpersuasive.\cite{endnote4}
To strengthen the argument we should set a robust upper limit
on $|(dS/dt)_{\rm life}|$, or equivalently on the total
entropy reduction $|\Delta S_{\rm life}|$, in a way that does not depend on
such an assumption.

\section{A robust argument}
\label{sec:myestimate}
Let us establish an upper limit on 
$|\Delta S_{\rm life}|$ 
by estimating the relevant
quantities in a way that is certain to overestimate the
final result.
Consider the entropy difference
between two systems: Earth as it
is at present, and a hypothetical Dead-Earth 
on which life never
evolved. We will assume that Dead-Earth and Earth
are identical, except that every atom in Earth's biomass is located in
Dead-Earth's atmosphere in its simplest molecular form.
When considering the entropy of Earth, we will assign zero
entropy to the biomass. That is, we will imagine that to
turn Dead-Earth into Earth, it is necessary to pluck every atom 
required for the biomass 
from the atmosphere and place it into its exact present-day quantum state. 
These assumptions
maximize the entropy of Dead-Earth and minimize that of Earth, so the
difference between the two entropies grossly overestimates the
required entropy reduction for the production of life in its present
form.

With these assumptions, we can use the standard thermodynamic result $\mu/T = -\partial S/\partial N$, 
which implies $\Delta S = -N\mu/T$, to estimate the entropy difference as
\beq
\Delta S_{\rm life}=S_{\mbox{\scriptsize
earth}}-S_{\mbox{\scriptsize dead-earth}}
\approx {N_b\mu\over T},
\eeq
where $N_b$ is the number of molecules required
to make up the biomass and $\mu$ is a typical
chemical potential for a molecule in the atmosphere of Dead-Earth.
If we use standard
relations for an ideal gas,\cite{schroeder} we find $\mu/kT\sim -10$, so that 
$\Delta S_{\rm life}<0$ as expected. 
We can 
obtain a value for $N_b$ from the estimate\cite{whitman} that the total
carbon biomass of Earth is
$\sim 10^{15}$\,kg. Even if we increase this value 
by a generous factor of 100
to account for other elements, we still have fewer
than $10^{43}$ molecules. 
We conclude that the entropy reduction required for life on
Earth is far less than
\beq
|\Delta S_{\rm life}|\sim 10^{44}k.
\eeq
If we compare this value with the rate of entropy production due to sunlight in
Eq.~(\ref{eq:ssun}), 
we find that the second law, in the form of Eq.~(\ref{eq:secondlaw}), 
is satisfied as long as the time required
for life to evolve on Earth is at least
\beq
\Delta t={|\Delta S_{\rm life}|\over (dS/dt)_{\rm sun}}\sim 10^7\,{\rm s},
\eeq
or less than a year.
Life on Earth took four billion years to evolve, so 
the second law of thermodynamics is safe.\cite{endnote5}

\begin{acknowledgments}
I thank Andrew Bunn, H.\ Franklin Bunn,
and Ovidiu Lipan for helpful discussions. Two
anonymous referees provided very useful comments that significantly
sharpened the arguments in this article. 
\end{acknowledgments}


\begin{thebibliography}{10}
\bibitem{endnote1}
The reader who wishes to compare this article with that of Ref.~\onlinecite{styer} may find it helpful to note that Ref.~\onlinecite{styer} calculates both $(dS/dt)_{\rm sun}$ and $(dS/dt)_{\rm CMB}$. In the quantitative conclusion [Eqs.~(5) and (6)], Ref.~\onlinecite{styer} correctly uses $(dS/dt)_{\rm sun}$, referring to this quantity as the ``entropy throughput.''
\bibitem{endnote3}Imagine that there are $n_a$ amino acids in solution, with $N$ available quantum states. Nondegeneracy means that $N\gg n_a$. The multiplicity is $\Omega(n_a)=\binom{N}{n_a}$. Taking one molecule out of solution causes the multiplicity to decrease by a factor $\Omega(n_a)/\Omega(n_a-1)=(N-n_a+1)/n_a\approx N/n_a=n_Q/n$.
\bibitem{endnote4}
If we adopt a much narrower view of which changes are due to evolution, these estimates would not apply. Even in this case, the factor of $10^3$ would still lack justification.
\bibitem{endnote5}
Creating all of life in six days might be thermodynamically problematic,
however.
\end{thebibliography}
\end{document}